\documentclass[floatfix,article,onecolumn]{revtex4}
\usepackage{graphicx}
\usepackage{xspace}
\usepackage{color}
\usepackage{amsmath}
\newcommand\beq{\begin{equation}}
\newcommand\bear{\begin{eqnarray}}
\newcommand\eeq{\end{equation}}
\newcommand\eear{\end{eqnarray}}

\begin{document}

\baselineskip=24pt

\title{Cyclopentadienyl-Benzene Based Sandwich Molecular Wires Showing Efficient Spin Filtering, Negative Differential Resistance, and Pressure Induced 
Electronic Transitions}

\author{\bf Prakash Parida$^{1}$,  Ershaad Ahamed Basheer$^1$  and Swapan K Pati$^{1,2}$}


\affiliation{$^{1}$Theoretical Sciences Unit and $^{2}$New Chemistry Unit\\
Jawaharlal Nehru Centre For Advanced Scientific Research \\
Jakkur Campus, Bangalore 560064, India \\ 
E-mail: pati@jncasr.ac.in}

\newcommand{\vant}{$\mathrm{[V_4(Ant)_2]_\infty}$}
\newcommand{\feant}{$\mathrm{[Fe_4(Ant)_2]_\infty}$}
\newcommand{\vfetopant}{$\mathrm{[V_2Fe_2(Ant)_2]_\infty}$}
\newcommand{\vfesideant}{$\mathrm{[(VFe)_2(Ant)_2]_\infty}$}
\newcommand{\vbzcp}{$\mathrm{[VBzVCp]_\infty}$} 
\newcommand{\febzcp}{$\mathrm{[FeBzFeCp]_\infty}$}
\newcommand{\febzvcp}{$\mathrm{[FeBzVCp]_\infty}$}
\newcommand{\be}{$\mathrm{E_b}$\xspace}
\newcommand{\bohrmag}{$\mathrm{\mu_B}$\xspace} 

\date{\today}
\vspace*{0.2cm}
\begin{abstract}
Using density functional theory, we investigate TM-cyclopentadienyl-benzene sandwich molecular wires (SMWs) which are composites of
TM-cyclopentadienyl and TM-benzene wires (TM = transition metal (V, Fe)). All the SMWs are found to be highly stable 
ferromagnetic half-metals, showing spin switching behavior. Transport calculations show that finite size clusters display spin filter property 
when coupled with Au electrodes on either side. 
$I-V_b$ characteristics of all systems confirm the spin filter property, with Au-BzVCpVBz-Au displaying
exceptionally high performance. In addition to spin filtering, the Au-BzFeCpFeBz-Au system also shows negative
differential resistance (NDR). Compression causes an
abrupt reduction in magnetic moment and a transition to a metallic phase, while stretching causes an increase in magnetic moment. 
Half-metallicity is preserved for modest amounts of stretching and compression.
\end{abstract}

\maketitle

\section{Introduction}
As conventional electronics approach the limits of miniaturization,\cite{ratner,ratner2,prinz,sanvito1,sanvito2,sharma,Cox} the focus of research is on developing devices that 
encode information using not only the charge of electrons but also their intrinsic spin. The long coherence time of spins and the 
ease of manipulating spins with external magnetic fields translates into devices that are more robust, smaller and more energy 
efficient than charge based devices. The essential building blocks of any spintronic device are a) A source of spin polarized 
current called a spin injector, b) a system that is sensitive to the presence of non-equilibrium spin called a spin detector, and c) a system 
that manipulates spins, either by application of external magnetic fields or by utilizing effective fields that arise out of spin orbit coupling in 
the substrate material. Spin manipulation can also be achieved through confinement geometries of the transport channel.                                                                              
Half-metals, a new class of compounds which exhibit large carrier spin polarizations, show promising features for spintronics 
applications, owing to the coexistence of the metallic nature of conduction in one spin orientation and an insulating nature for the 
other spin orientation.\cite{Groot,Katsnelson} Many theoreticians and experimentalists are enthusiastically pursuing the search for 
stable materials which have half-metallic and spin filter properties and thus can have huge potential in 
advanced spintronic devices.\cite{Wang,Cama,Fer,Oeiras,Sugaw,Mananes,Chioncel,Wang5,Mas}

Organometallic sandwich complexes are fast emerging as promising candidates for the role of a source of spin polarized current. The 
discovery of ferrocene $\mathrm{FeCp_2}$ in 1951\cite{kealy} initiated interest in the field of organometallic compounds. 
Further work resulted in the synthesis of several derivatives of $\mathrm{FeCp_2}$.\cite{wilkinson2} With the development of 
gas-phase synthesis of organometallic compounds with the aid of laser vapourization, a major impediment to the synthesis of multidecker 
sandwich complexes had been overcome. This technique has been used in the synthesis and experimental investigation of several complexes 
including transition metal-benzene ($\mathrm{TM_mBz_n}$) complexes, TM-cyclopentadienyl ($\mathrm{M_nCp_{n+1}}$) complexes and lanthanide 
rare earth metal-$\mathrm{C_8H_8}$ complexes.\cite{weis,kurikawa,nakajima,hoshino,miyajima3,miyajima4,kurikawa2,miyajima5,takegami,Jacob,Yasuike} Other 
successes in synthesis of organometallic sandwich complexes include the 
production of vanadium-iron-cyclopentadienyl ($\mathrm{V_n(FeCp_2)_{n+1}}$) by the reaction of V vapour with 
ferrocene ($\mathrm{FeCp_2}$),\cite{nagao} and vanadium-anthracene clusters by reacting V vapour 
with anthracene.\cite{kurikawa} Many theoretical studies predict half-metallicity in such 1-D organometallic 
sandwich molecular wires (SMWs), like $\mathrm{[VCp]_\infty}$, $\mathrm{[VBz]_\infty}$, $\mathrm{[V_2Anthracene]_\infty}$ 
and many others.\cite{pandey,pandey2,wang2,wang3,kua,weng,xu,sanvito3,
zhang3,li2,zhou,shen,liu1,liu2,parida1,parida2,zhu,maslyuk,kandalam,wu,lu,kang,wang,Xiang,yi,kol}  
However, it was found that during adsorption onto a 
self-assembled monolayer (SAM) matrix, these clusters dissociate into smaller fragments, which has raised questions about their stability.\cite{mitsui,nagaoka} A 
theoretical method has been proposed to 
overcome these experimental limitations, which suggests the use of a rigid scaffold as a support, to the clusters.\cite{mallajosyula,mallajosyula2}

\begin{figure}
\begin{center}
\includegraphics[width=1.0\textwidth]{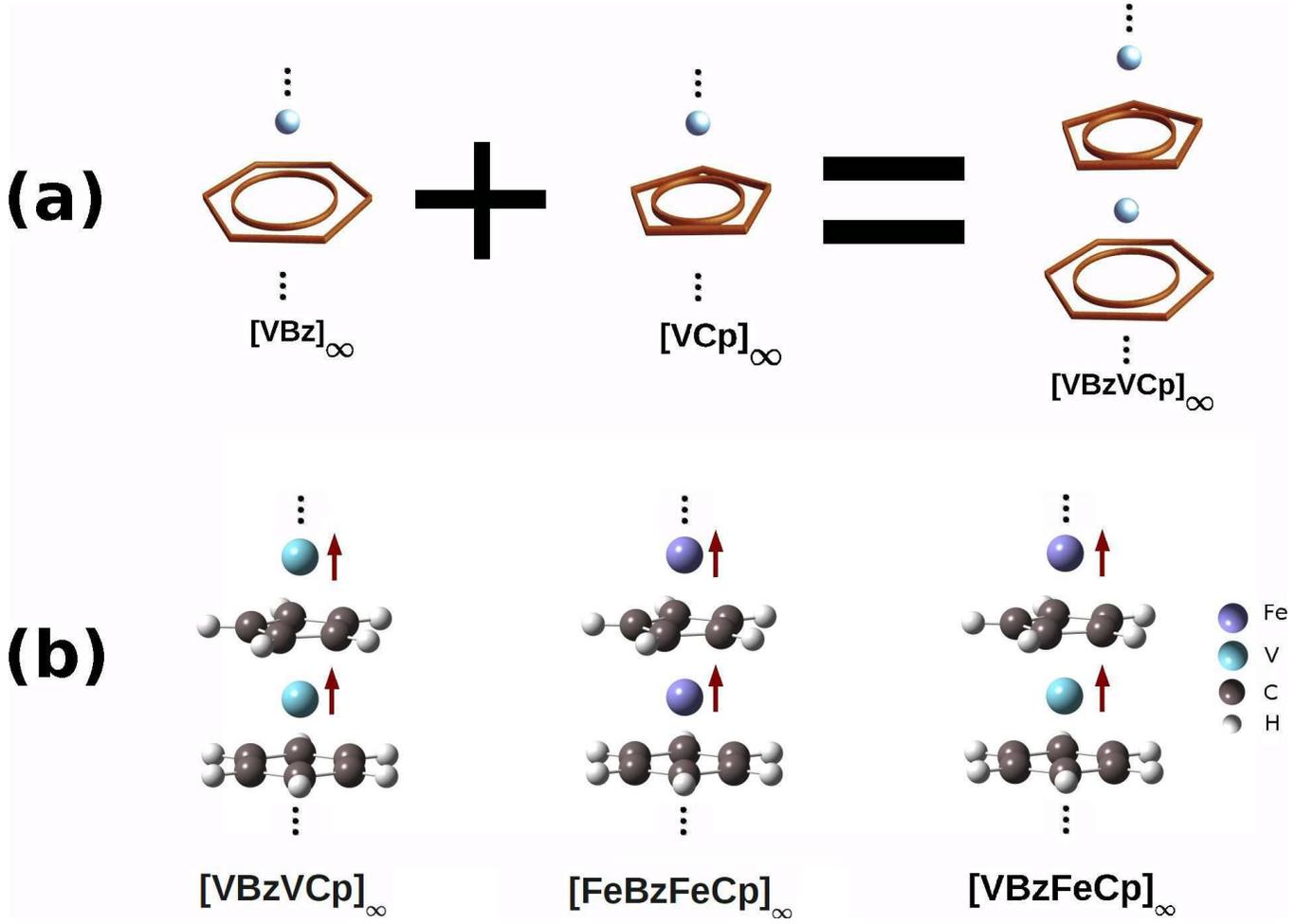}
\end{center}
\caption{(a) Schematic of V-cyclopentadienyl-benzene wire as a composite of V-benzene and V-cyclopentadienyl wires. (b) Optimized 
geometries with ground state spin configurations of the SMWs.}
\label{Fig1}
\end{figure}

  Since $\mathrm{[VCp]_\infty}$ is a strong half-metal (HM), and $\mathrm{[VBz]_\infty}$ is a quasi half-metal, it is interesting to ask following questions: 
(1) what will be the details of electronic structure of a composite system, which is composed of these two wires? (2) Will the composite wire show 
ferromagnetism and half-metallicity? (3) Can this composite system show some novelty in transport phenomenon? (4) What is the effect of pressure on this 
composite wire? In this article, with the aim of obtaining a thorough understanding of
these SMWs, we explore in detail the electronic, magnetic, and transport properties of these SMWs and also demonstrate
the effects of pressure on these properties. We model the unit cell of the 
composite wire, vanadium-cyclopentadienyl-benzene ($\mathrm{[VBzVCp]_\infty}$)  by stacking the unit 
cell of $\mathrm{[VBz]_\infty}$ and $\mathrm{[VCp]_\infty}$ on top of each other in the periodic direction 
(See schematic in Figure \ref{Fig1}). Our theoretical model is motivated by the early experimental synthesis of 
finite size $\mathrm{Cp_2V_2Bz}$ clusters.\cite{duff} For a detailed analysis, we also extend our study to two 
other SMWs, $\mathrm{[FeBzFeCp]_\infty}$ and $\mathrm{[VBzFeCp]_\infty}$. 
  
\section{Methods and Models}
The geometry optimization and electronic structure are calculated by using spin-unrestricted density functional theory (DFT) 
as implemented in the SIESTA\cite{siesta} DFT package. A double-$\zeta$ polarized (DZP) basis set is used with a real space mesh 
cutoff of 500 Ry. Exchange-correlation within the DFT formalism are calculated using the Perdew-Burke-Ernzerhof\cite{pbe} 
functional within the Generalized Gradient Approximation (GGA). 
To verify the robustness of our results obtained using localized basis and GGA functional within SIESTA package,
we have carried out few calculations using plane wave basis within
PWscf package of the Quantum-ESPRESSO distribution,\cite{pwscf} and also hybrid B3LYP functional\cite{b3lyp,b3lyp1} as implemented in CRYSTAL06 code.\cite{crystal} 
To verify the stability of finite size clusters, we have calculated their harmonic vibrational frequencies using Gaussian 03 program package.\cite{g03} 
The Monkhorst-Pack K sampling grid is set to 30 points 
in the periodic direction. The conjugate gradient method is used for geometry optimization. Vacuum separations of 
30\AA\mbox{} are used to suppress spurious interactions in the non-periodic directions. 
Our spin transport calculations are based on the non-equilibrium
Green’s function formalism (NEGF) as implemented in the TranSIESTA package,\cite{transiesta} extended to spin-polarised systems.
Here, we focus on electron transmission close to the
Fermi energy. The transmission is obtained from the equation
\begin{eqnarray}
T(E)= Tr[\Gamma_L (E) G^a(E)\Gamma_R(E)G^r(E)],
\end{eqnarray}
\noindent
where the retarded Green's function, $G^r(E)$ is calculated from the Hamiltonian and self-energies of the central region.
$\Gamma_{\alpha}(E)$ is (-2 times) the imaginary part of the self-energies of the left and right electrodes ($\alpha$=L,R).

\section{Results and Discussion}
\subsection{Stability and Electronic Properties} 
In the followings, all our discussions are mainly based on the results obtained using SIESTA package unless and otherwise it is mentioned.
In Figure \ref{Fig1}, we present the optimized geometries of $\mathrm{[VBzVCp]_\infty}$, $\mathrm{[FeBzFeCp]_\infty}$ and 
$\mathrm{[VBzFeCp]_\infty}$ SMWs. In each structure, the metal atom has $\eta^5$ and $\eta^6$ types of coordination with Cp and Bz 
rings on opposite sides. Although both V and Fe atoms have the same type of coordination, we find the interplanar separation 
between the Bz and Cp ring in the case of $\mathrm{[FeBzFeCp]_\infty}$ is less than that in $\mathrm{[VBzVCp]_\infty}$ (See the Table \ref{magtable}). 
Further, in order to estimate the stability of the SMWs, we calculate the 
binding energies (\be) of their respective unit cells. We define the binding energies of these SMWs as 

\begin{align}
\mathrm{E_b([VBzVCp]_\infty)}&= \mathrm{E([VBzVCp]_\infty)- 2E(V) - E(Bz) - E(Cp)}\\
\mathrm{E_b([FeBzFeCp]_\infty)} &= \mathrm{E([FeBzFeCp]_\infty)- 2E(Fe) - E(Bz) - E(Cp)}\\
\mathrm{E_b([VBzFeCp]_\infty)} &= \mathrm{E([VBzFeCp]_\infty)- E(V)-  E(Fe) - E(Bz) - E(Cp)}
\end{align}

Since SIESTA uses atomic orbital basis, it is essential to account for basis set superposition error (BSSE)\cite{bsse} for energetics of each sandwich complex.  
It is found that BSSE corrections are significant for all SMWs (see Table \ref{magtable}).    
As can be seen from Table \ref{magtable}, all the SMWs are highly stabilized in comparison to their isolated units.  
\begin{table}
\begin{center}
\begin{tabular}{|c|c|c|c|c|c|c|c|}
\hline
\rule{0pt}{2.5ex} SMW & $c(\mathrm{\AA})$ & \be(eV) & BSSE (eV) & MM(\bohrmag) & $\Delta E\mathrm{(meV)}$ & $\Delta_{maj}\mathrm{(eV)}$ & $\Delta_{min}\mathrm{(eV)}$ \\
\hline
$\mathrm{[VBzVCp]_\infty}$   & 3.6  & -14.21 & 0.73 & 3.0 & -287  & 0.98      &  metallic \\
$\mathrm{[FeBzFeCp]_\infty}$ & 3.41 & -11.30 & 1.57 & 3.0 & -364   & metallic  & 1.05\\
$\mathrm{[VBzFeCp]_\infty}$ & 3.54 & -12.97 &  1.25  &  4.0 &  -320  & 2.72      & metallic \\
\hline
\end{tabular}
\end{center}
\caption{Optimized Bz-Cp interlayer separation (c), binding energy (\be), BSSE correction , total magnetic moment per unit cell (MM), Energy difference 
between ferromagnetic and antiferromagnetic states ($\Delta E$), the band gap for majority ($\Delta_{maj}$) and minority spin ($\Delta_{min}$) of examined 1D SMWs.}
\label{magtable}
\end{table}

\begin{figure}
  \begin{center}
   \includegraphics[width=1.0\textwidth]{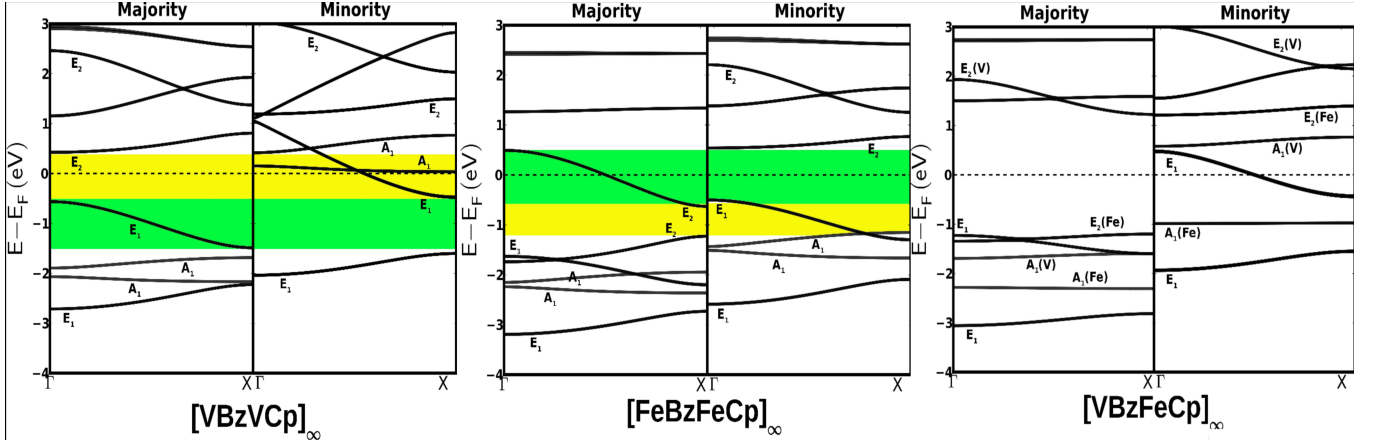}
  \end{center}
  \caption{Spin polarized band structure for the SMWs. Fermi level is scaled to zero.
Green and yellow shaded regions show spin switching behavior near Fermi energy. Majority spin states are
allowed in green regions while minority spin states are allowed in yellow regions.}
        \label{Fig2}
\end{figure}

To probe magnetic interaction in these SMWs, we calculate the magnetic moment on the unit cells of each SMW. We 
find all the systems to be ferromagnetically stabilized and thus have a magnetically polarized state irrespective of 
the basis set (localized or plane wave) and functionals (GGA or B3LYP) used.   
In order to estimate the ferromagnetic stability, we calculate the energy difference ($\Delta E$) between the 
ferromagnetic (FM) and the anti-ferromagnetic (AFM) state. Note that, the negative sign of $\Delta E$ indicates the 
stabilization of the FM state over the AFM state. The large magnitude of $\Delta E$ (See Table. \ref{magtable}) indicates 
the high stability of the FM state even at room temperature. From our calculations, we find that the net magnetic moment of 
the unit cell is 3\bohrmag for $\mathrm{[VBzVCp]_\infty}$ and $\mathrm{[FeBzFeCp]_\infty}$, and 4\bohrmag for 
$\mathrm{[VBzFeCp]_\infty}$ SMWs (see Table \ref{magtable}), and this value almost remain same for the plane wave basis and B3LYP functional. A 
local magnetic moment of 1.7\bohrmag on V and and 1.6\bohrmag on Fe atom are found in $\mathrm{[VBzVCp]_\infty}$ and 
$\mathrm{[FeBzFeCp]_\infty}$ wires, respectively. In fact, while 
two V (or Fe) atoms contribute a positive magnetic moment of +3.3\bohrmag, Bz and Cp rings together contribute a small 
negative moment of -0.3\bohrmag to the total magnetic moment (3\bohrmag) of the unit cell. While for SMW of mixed 
metal, $\mathrm{[VBzFeCp]_\infty}$, the net magnetic moment of the unit cell is 4\bohrmag, with a local magnetic 
moment of +1.9\bohrmag on V and +2.4\bohrmag on Fe atoms. In order to obtain a better picture of electronic structure and 
the origin of magnetic moment, we display spin polarized band 
structures in Figure \ref{Fig2}. Under $D_{5h}$ and/or $D_{6h}$ symmetry of crystal field, the 3d orbitals of Fe and 
V split into a singlet $A_1$ ($d_{z^2}$), and two sets of doublets $E_1 (d_{xy}, d_{x^2-y^2})$ and $E_2 ( d_{xz}, d_{yz} )$. 
Furthermore, we note that, although the isolated V atom has the valence electronic configuration $3d^34s^2$, due to strong hybridization, the vanadium 4s 
levels are shifted above $E_F$ making the effective valence configuration $3d^54s^0$. For $\mathrm{[TM-Cp]_\infty}$ SMWs (TM = Transition Metal), Shen et al. have 
proposed a general mechanism which involves transfer of one electron from the metal atom to the Cp ring.\cite{shen} 
We utilize the same mechanism to explain the band structure and origin of magnetic moment. Note that, there are two d-derived bands 
for each of five d-orbitals. This is because there are two metal atoms in the primitive cell of our SMWs. Hence, there are always 
two bands for each band of $A_1$, $E_1$ or $E_2$ symmetry. In the rest of our discussion, when we mention two bands of $E_1$ (or $E_2$) symmetry, it is 
implied that it consists of 4 bands.

\begin{figure}
\begin{center}
\includegraphics[width=1.0\textwidth]{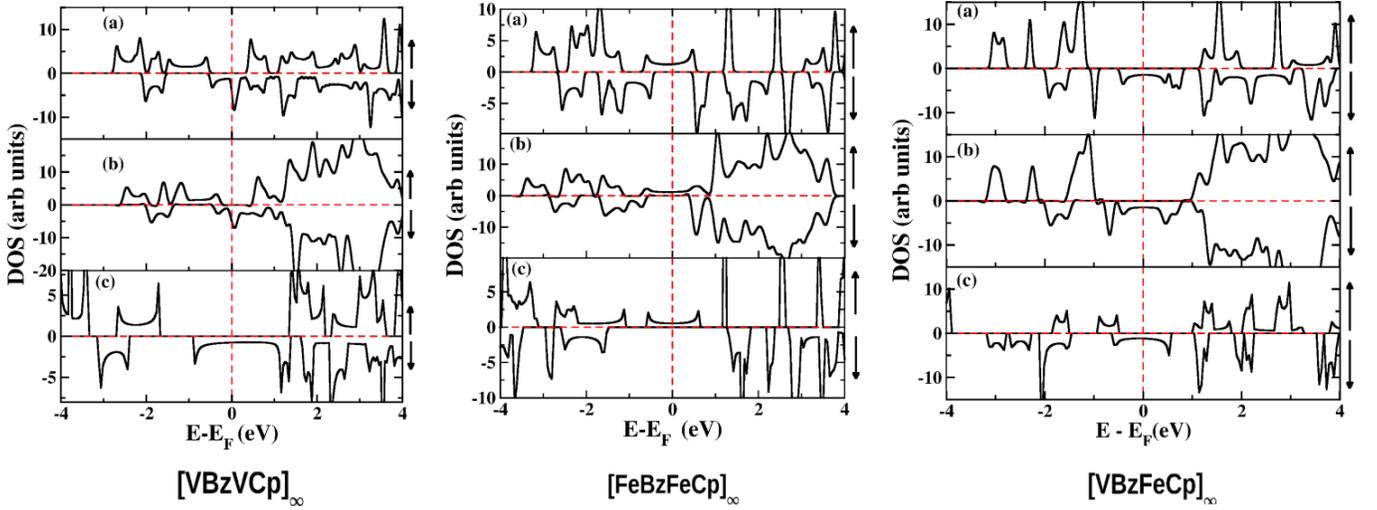}
\end{center}
\caption{Spin polarised DOS plot for the SMWs. Top panel (a) shows the results for localised basis (SIESTA), middile panel shows for plane wave basis (Quantum espresso)
and bottom panel shows for B3LYP functional (CRYSTAL06). Up-arrow and down-arrow show the results for majority spins and minority spins, respectively.}
        \label{Fig3}
\end{figure}

The unit cell of $\mathrm{[VBzVCp]_\infty}$ contains two V atoms, i.e., ten d-electrons. Following the transfer of 1 electron from 
V atoms to the Cp ligand, the total number of d-electrons becomes 9. From the band structure, it is clear that out of these 
9 d-electrons, 6 majority spin electrons completely fill two bands of $A_1$ and two bands of $E_1$ symmetry, leading to 
the opening of an semiconducting gap of 0.98 eV at the $\Gamma$ point in the majority channel. Out of the remaining 
3 d-electrons, 2 electrons completely fill one of the two minority bands which have $E_1$ symmetry, the other minority 
band being partially filled by the remaining 1 electron. In fact, due to such partial filling, one 
of the $E_1$ bands crosses $E_F$ in the minority spin channel, leading to metallic behavior in the 
electronic structure for minority spins. The important point to note here is that, we have a coexistence 
of the metallic and semiconducting nature for electrons in the minority and majority spin 
channels, respectively, which leads to half-metallic behavior for the $\mathrm{[VBzVCp]_\infty}$ SMW. It is 
clear from the band structure that, two localized $A_1$ bands, which are completely filled for
majority spin and completely empty for the minority spin channel, contribute magnetic 
moment of 2\bohrmag and the partially filled $E_1$ band contributes 1\bohrmag to the total 
magnetic moment of 3\bohrmag. We would like to point out that, due to strong 
hybridization, the states with the $E_2$ symmetry move far away from the Fermi level and remain 
silent in transport and magnetic behaviors. Very interestingly, $\mathrm{[VBzVCp]_\infty}$ shows 
a nice spin switching behavior around Fermi energy. 
As it is clear from Figure \ref{Fig2} (green and yellow shaded regions), only majority spin states are allowed in the energy window from -1.5 eV to -0.6 eV 
while only minority spin states are favored in the energy window from -0.6 to 1.1 eV, showing a clear spin switching behavior. 
Note that, this kind of spin switching behavior is not        
noticed in the band structure of $\mathrm{[VBz]_\infty}$ and $\mathrm{[VCp]_\infty}$.

  Similarly, for $\mathrm{[FeBzFeCp]_\infty}$, the effective electronic configuration for Fe is $3d^84s^0$. As 
the SMW contains two Fe atoms and one of the d-electrons is transferred from Fe, 15 d-electrons are available for 
filling the bands. Out of these 15 d-electrons, 6 minority electrons completely fill two bands of $E_1$ symmetry and 
two bands of $A_1$ symmetry, leading to the opening of a semiconducting gap of 1.05 eV at the $\Gamma$ point in the 
minority spin channel. While out of the remaining 9 electrons, 8 majority electrons completely fill two bands of 
each from $E_1$ and $A_1$ symmetry and one of the bands of $E_2$ symmetry. The remaining 1 majority spin electron 
partially fills the other band of $E_2$ symmetry and hence crosses the Fermi level, leading to metallic behavior in 
the majority spin channel. Thus, $\mathrm{[FeBzFeCp]_\infty}$ wire is a perfect ferromagnetic half-metal. One of the 
$E_2$ bands which is completely filled for majority spin and completely empty for minority spin electrons, contributes 
a magnetic moment of 2\bohrmag and the other partially filled $E_2$ band contributes 1\bohrmag to the total magnetic 
moment (3\bohrmag) of the system. We wish to emphasize here that, the orbitals of $E_1$ ($d_{xy}, d_{x^2-y^2}$) and $ A_1$ ($d_z^2$) 
symmetry play an important role in the magnetic interaction in $\mathrm{[VBzVCp]_\infty}$, while for the case of 
$\mathrm{[FeBzFeCp]_\infty}$, the orbitals of $E_2$ symmetry ($d_{xz}, d_{yz}$) are important for magnetism.
Interestingly, $\mathrm{[FeBzFeCp]_\infty}$ also show spin switching behavior unlike $\mathrm{[FeBz]_\infty}$ and $\mathrm{[FeCp]_\infty}$
(see the shaded regions in Figure \ref{Fig2}).

 For the mixed metal SMW, $\mathrm{[VBzFeCp]_\infty}$, out of the total 12 d-electrons, 8 majority spin electrons completely 
fill two bands of $A_1$ (both V and Fe contribute), two bands of $E_1$ and one band of $E_2$ (derived from Fe). 
Four minority spin electrons fill completely one $A_1$ (derived from Fe only) band, one of the $E_1$ and 
partially the other $E_1$ band. Further, we find that one of the $A_1$ bands (derived from V) contributes 
1\bohrmag, $E_2$ (derived mostly from Fe) contributes 2\bohrmag and one of the $E_1$ contributes 1\bohrmag, to the total 
moment of 4\bohrmag in the unit cell. A wide gap of 2.72 eV opens up at the zone boundary (X point) in the majority spin 
channel, while due to partial filling, bands of $E_1$ symmetry cross the Fermi level in the minority spin channel, making the 
system a strong half-metal. In fact, we would like to stress that, due to the large magnetic moment, the $\mathrm{[VBzFeCp]_\infty}$ system 
would be much more useful for molecular magnets. 

 In order to verify the results indicating half-metallicity that were obtained using localized basis and GGA functional,
we perform additional calculations for all the systems within plane wave basis as well as within the hybrid B3LYP functional. 
We present the DOS plots for the different levels of calculations for all the systems in Figure \ref{Fig3}. 
As can be seen, at the Fermi level a gap opens for one spin channel, while there is a finite density of states for the other spin channel, leading 
to the half-metallicity in all the systems. All the systems retain their half-metallicity irrespective of the level of the calculations. 
The size of the gap and the relative positions of energy levels are of course subject to change with the method chosen,
but the half-metallic behavior is robust against details of the calculations.
 
\begin{table}
\begin{center}
\begin{tabular}{|c|c|c|c|c|c|c|}
\hline
\rule{0pt}{2.5ex}Cluster & d ($\AA$) & BE (eV) & $T_S$(\%) & $HLG^\uparrow$ (eV) & $HLG^\downarrow$ (eV) & $|HLG^\uparrow - HLG^\downarrow|$ (eV)\\
\hline
BzVcpVBz & 2.61 & -171.3 & 94 & 2.20 & 0.50 & 1.7\\
BzFeCpFeBz & 2.78 & -172.8 & 70  & 0.20 & 1.65 & 1.45\\
BzVCpFeBz &  2.69 & -172.1 & 58 & 1.98 & 0.68 & 1.30\\
\hline
\end{tabular}
\end{center}
\caption{Distance (d) between finite size cluster and gold electrode, binding energy (BE) of cluster-electrode system,
spin polarization of transmission $T_S = |T_{min}(E_F)-T_{maj}(E_F)| / [T_{min} (E_F) + T_{maj} (E_F)]$,
HOMO-LUMO gap for majority spin ($HLG^\uparrow$), HOMO-LUMO gap for minority spin ($HLG^\downarrow$) and the difference
between $HLG^\uparrow$ and $HLG^\downarrow$ for each system.}
\label{transtable}
\end{table}

\subsection{Transport Properties} 
In order to explore the possibility of real world application in devices, we investigate the transport properties of 
these finite size clusters. Three finite size clusters, BzVCpVBz, BzFeCpFeBz and BzVCpFeBz, coupled with gold electrodes on 
either side are modeled. All the finite size clusters are optimized and then the structural stabilities are verified by 
the calculation of their harmonic vibrational frequencies.  
We consider a nonmagnetic 3x3x3 bulk gold electrode (Au (111) plane) consisting of 27 Au atoms for transport calculations.  
Following the calculation of optimized geometries for each of the isolated finite size clusters, we calculate the optimal separation 
between a single electrode and the finite cluster. Using this optimal separation, we place electrodes symmetrically on 
either end of the finite cluster and optimize the finite cluster geometry while freezing the coordinates of the electrodes. 
A periodic boundary condition is applied along the transverse direction in scattering region (electrode-cluster-electrode) which includes
3 layers of gold. For the device structure and stability, we have calculated binding energy of the electrode-cluster systems  
and relaxed distance (d) between the cluster and the gold electrode (see Table \ref{transtable}). 
All the clusters are found to be adsorbed strongly to the gold surface, with BzFeCpFeBz shows maximum binding strength,
while the distance between the benzene ring and the gold electrode is less in case of BzVCpVBz.
 
\begin{figure}
  \begin{center}
  \includegraphics[width=1.0\textwidth]{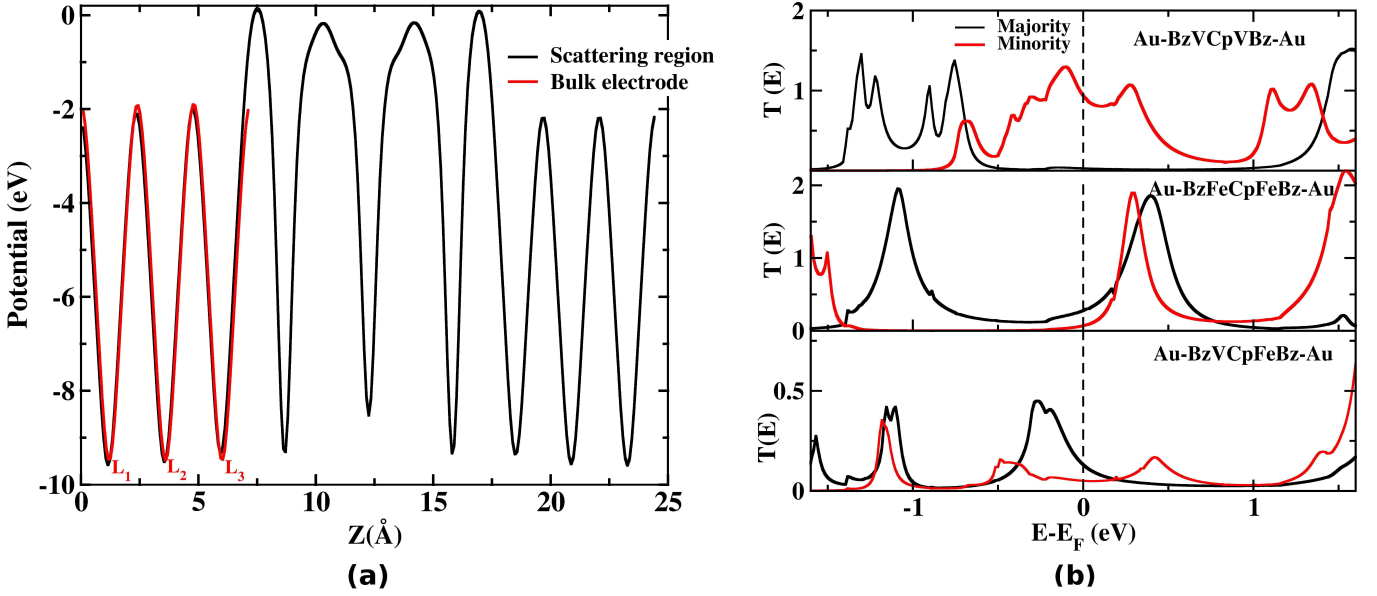}
  \end{center}
  \caption{(a) Electrostatic potential (averaged over xy plane) as a function of Z for Au-BzVCpVBz-Au. $L_1, L_2$ and $L_3$
represent the minimas in potential profile at the position of each layer of 3x3x3 gold electrode, (b) spin polarized zero-bias
transmission function, T(E) for each of the electrode-cluster-electrode systems.}
        \label{Fig4}
\end{figure}

\begin{figure}
\begin{center}
\includegraphics[width=1.0\textwidth]{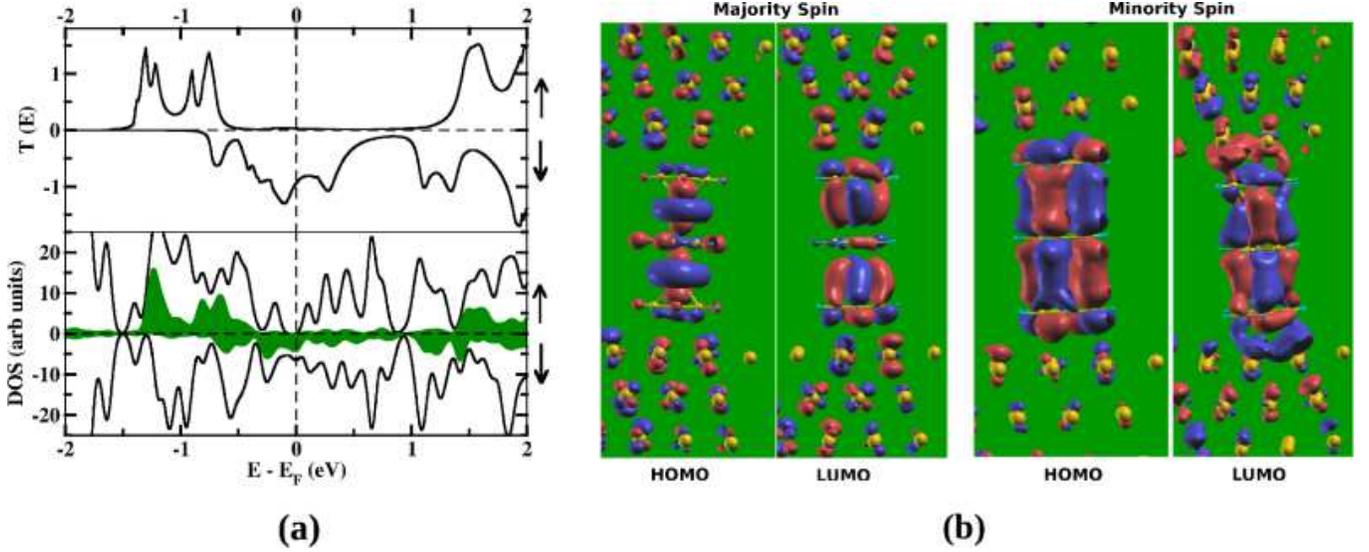}
\end{center}
\caption{(a) Top and bottom panel shows spin-polarized zero-bias transmission and DOS plot for Au-BzVCpVBz-Au.The shaded green region shows DOS
projected onto the molecular state of BzVCpVBz. (b) Orbital plots of HOMO and LUMO of Au-BzVCpVBz-Au. 
Up-arrow and down-arrow show the results for majority spins and minority spins, respectively.}
\label{Fig5}
\end{figure}

In Figure \ref{Fig4}(a), we plot electrostatic potential (averaged over xy plane) as a function of Z for both scattering region and
bulk electrode in the case of Au-BzVCpVBz-Au. We find that the potential at the border of scattering region
approaches quite well to the bulk values of gold electrode, which confirms that our inclusion of only three layers of gold
into the scattering region is good enough to have less scattering at the interface.
In Figure \ref{Fig4}(b) we present the spin polarized zero-bias transmission function T(E) for different systems.
The spin polarization of transmission is defined as 
\begin{equation}
T_S = \frac{|T_{min}(E_F)-T_{maj}(E_F)|}{[T_{min} (E_F) + T_{maj} (E_F)]},
\end{equation}
where $T_{min}$ and $T_{maj}$ represent the transmission coefficient of minority and majority spin channels, respectively. From 
the Table \ref{transtable}, it can be seen that these systems show high spin polarization ($T_S$). 
The order of the $T_S$ value in different systems when coupled with gold electrodes is : BzVCpVBz $>$ BzFeCpFeBz $>$ BzVCpFeBz (see Table \ref{transtable}).

\begin{figure}
\begin{center}
\includegraphics[width=1.0\textwidth]{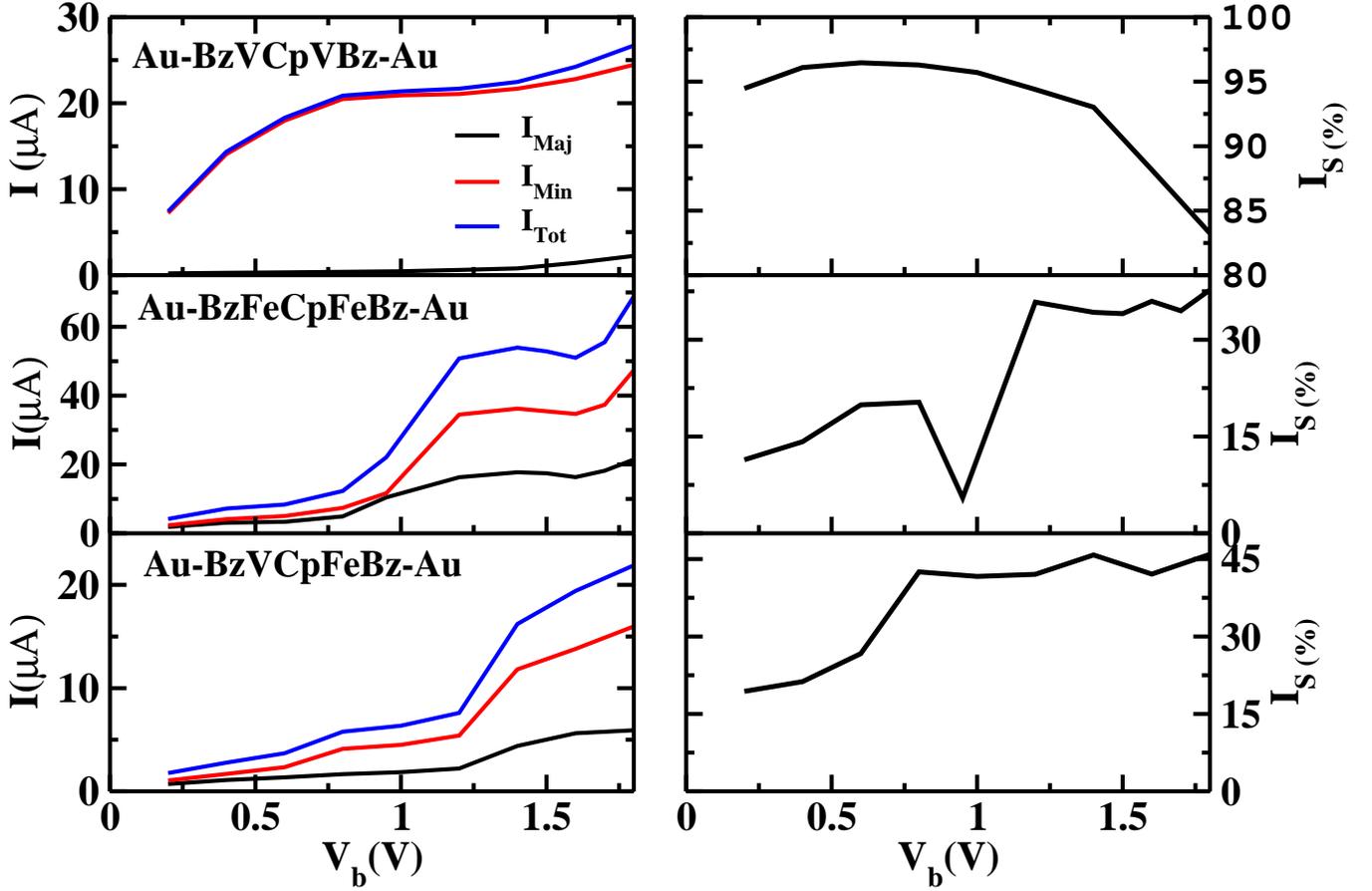}
\end{center}
\caption{ Left panel shows $I-V_b$ characteristics. Black, red and blue lines represent $I_{Maj}$, $I_{Min}$ and $I_{Tot}(I_{min}+I_{maj})$.  
Right panel shows spin filter efficiency ($I_S$) for different systems. Top, middle and bottom panel shows the results for
Au-BzVCpVBz-Au, Au-BzFeCpFeBz-Au and Au-BzVCpFeBz-Au, respectively.}
\label{Fig6}
\end{figure}

To understand this order, we calculate the energy levels of highest occupied molecular orbital (HOMO) and lowest unoccupied molecular orbital (LUMO) 
for isolated molecules. Table \ref{transtable} lists the gaps between the HOMO and the LUMO in the majority and minority spin 
channels ($HLG^\uparrow$ and $HLG^\downarrow$). It is clear that larger the difference in the value of 
the HOMO-LUMO gap between both the spin channel ($HLG^\uparrow$ - $HLG^\downarrow$), the larger is the $T_S$ value.
To understand the high spin filter efficiency of Au-BzVCpVBz-Au, in Figure \ref{Fig5} (a) we plot the total DOS of electrode-molecule-electrode (EME) system and 
its projection (pDOS) onto the molecular states. 
As can been seen, the transmission spectra shows a series of peaks with strong correlation between transmission and 
PDOS spectra, especially with regard to the location of their peaks.
These transmission peaks correspond to the resonant transmission through molecular states.
Transmission shows a peak only when molecular states resonate with the states of the electrodes.
In a low energy window around the Fermi energy, the vanishing contribution of molecular states to 
the eigenstates of the system leads to a case of weak resonance, which in turn makes the transmission
zero (negligible) for majority spin electrons. While due to strong resonance, minority spin electrons show 
strong transmission peaks near the Fermi level. Looking once again at the plots of HOMO and LUMO for majority 
and minority spin electrons in Figure \ref{Fig5} (b), 
it can be seen that due to weak coupling of molecular orbitals and incident states from the electrode, the HOMO and 
LUMO rarely contribute to the transmission for majority spin. While the HOMO and LUMO for the minority 
spin show a strong transmission peak because of the strong resonance of molecular states with the electrode. 

\begin{figure}
\begin{center}
\includegraphics[width=1.0\textwidth]{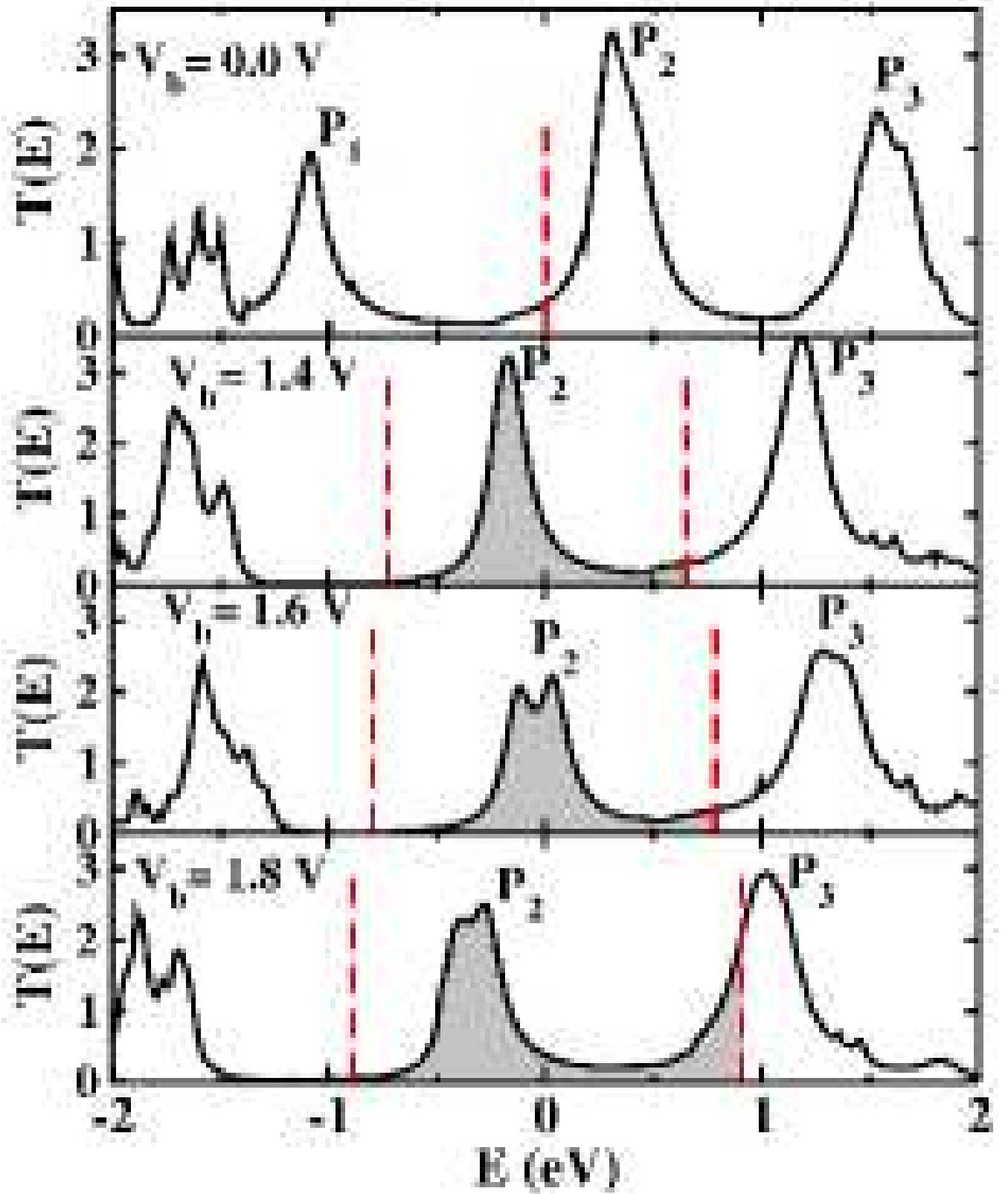}
\end{center}
\caption{ Total Transmission function at different bias voltage for Au-BzFeCpFeBz-Au . Red dashed line represent the bias window
          and the shaded region shows the integral area in the bias window.}
        \label{Fig7}
\end{figure}

  As actual devices work at a finite bias, in Figure \ref{Fig6} we present $I-V_b$ characteristics.
The spin filtering effect is clearly illustrated by the $I-V_b$ plot.
Particularly, in the case of 
Au-BzVCpVBz-Au, when bias is applied, the current reaches up to several $\mu$A for the minority spin component, however
remains almost zero for the majority spin component. To quantify it, we define spin polarization at finite bias in terms of 
spin-resolved currents 
\begin{equation*}
I_S = \frac{|I_{min}-I_{maj}|}{(I_{min}+I_{maj})},
\end{equation*}
where $I_{min}$ and $I_{maj}$ represent minority and majority currents, respectively.
It is quite clear that, all the systems show efficient spin filtering, while Au-BzVCpVBz-Au notably shows
a very high value of $I_S$ = 96.5\% at a bias voltage, $V_b$ = 0.6 V. Besides the spin filtering, an interesting   
negative differential resistance (NDR)\cite{parida3,parida4,natalya,nikita} effect is also observed in the bias range of 1.4 - 1.6 V for Au-BzFeCpFeBz-Au system.
To understand the NDR behavior, we plot the total transmission function T(E, V$_b$) of Au-BzFeCpFeBz-Au system
at various bias in Figure \ref{Fig7}. 
The current is calculated by the Landauer-Buttiker formula $I = (e/h)\int ^ {\mu_R} _ { \mu_L } T (E,V_b) dE$.
The transmission coefficient $T(E,V_b)$ is the function of the energy (E) and bias window $V_b$ 
(integral window). The current thus depends upon the integral area (namely, the shaded area in Figure \ref{Fig7}) which
is related to two factors: one is the transport coefficient and the other is the magnitude of the bias window. 
In Figure \ref{Fig7}, we mark the three significant peaks ($ P_1, P_2, P_3 $) in $T(E,V_b=0)$. As the bias increases
these peaks gradually shift toward the low-energy side, accompanied by a change in height. When $V_b$ reaches 
1.4 V, the peak $P_2$ drifts into the bias window, which in turn increases the available integral area, resulting in a 
significant rise in the electric current. When $V_b$ further increases to 1.6 V, the peak value of $P_2$ reduces,
reducing available integral area and results in a drop in current, which causes NDR.        
With a further increase in $V_b$, $P_3$ peak enters into the bias window at $V_b$ = 1.8 V, causing 
a significant rise in the current.    

\begin{figure}
\begin{center}
\includegraphics[width=1.0\textwidth]{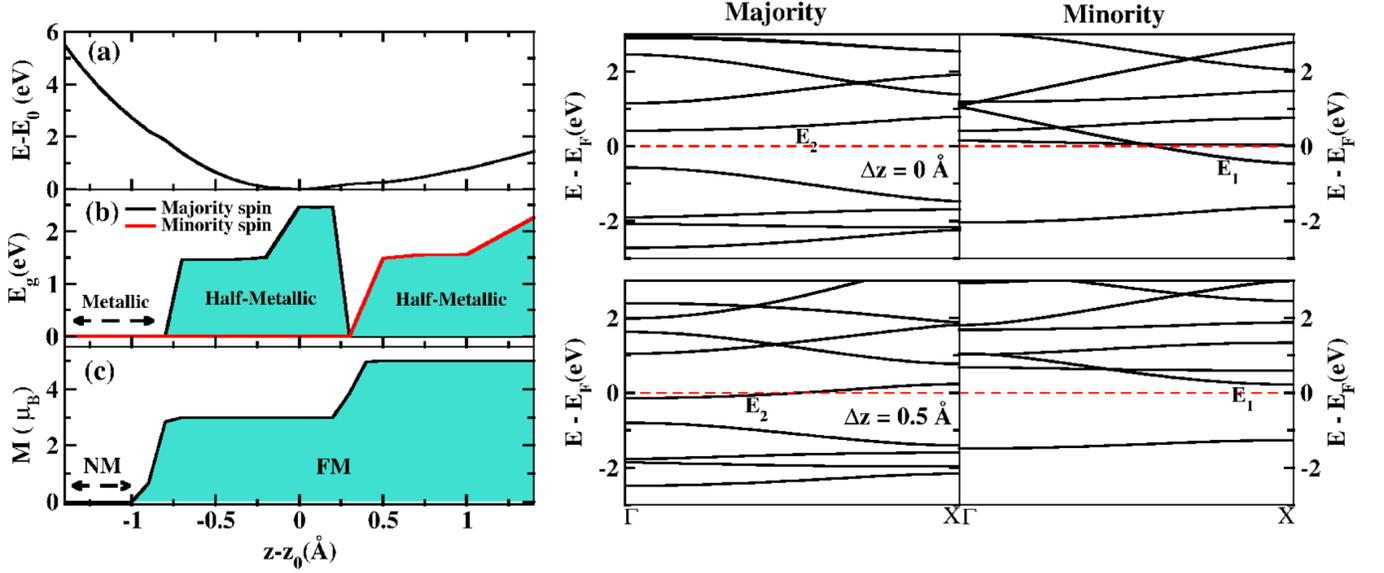}
\end{center}
\caption{ Left panel: (a) shows the energy (E) of $\mathrm{[VBzFeCp]_\infty}$ scaled to the energy of its equilibrium structure ($E_0$)
as a function of change in lattice constant ($z-z_0$). (b) shows evolution of spin-polarized band gap($E_g$) and (c) shows
variation of magnetization under stretching and compression of the $\mathrm{[VBzFeCp]_\infty}$ wire.
Right panel: Spin-polarized band structure for $\mathrm{[VBzVCp]_\infty}$ for the case of $\Delta$z = 0 \AA ~ and $\Delta$z = 0.5 \AA.}
\label{Fig8}
\end{figure}

\subsection{Pressure Induced Transitions}     
Pressure is well known to have an effect on the electronic and magnetic properties of magnetic materials.  
Pressure can imply either compression or stretching, which we model by changing the lattice constant. 
As \vbzcp wire shows strong half-metallicity and its finite size cluster shows efficient spin-filter property,  
we choose \vbzcp wire as a case study to understand the effect of pressure. In Figure \ref{Fig8}, we show the the energy of 
\vbzcp scaled to the corresponding energy at equilibrium lattice constant ($z_0=7.2\AA$)
together with band gap and magnetic moment as a function of the lattice constant. 
We find that, with the application of pressure, the AFM state is favored less, as compared to FM or non-magnetic (NM) states. 
While subject to stretching, the majority spin gap initially remains constant, and then suddenly drops to zero 
at a critical value of $\Delta$z = z-$z_0$ = 0.3 \AA. As the  $\Delta$z increases beyond this point, the minority spin gap begins to 
open.  This is because, on stretching, the partially occupied minority spin $E_1$ band is pushed above $E_F$ as a fractional charge 
flows from this band to the upper lying majority spin band of $E_2$ symmetry which in turn passes through the Fermi level
(See Figure \ref{Fig8}: right panel). Thus, the transition is accompanied by an abrupt jump in the magnetic moment 
from 3 \bohrmag to 3.8 \bohrmag and then to 5 \bohrmag.
Under compression, the majority spin gap gradually reduces and abruptly drops to zero at a 
critical point, $\Delta$z = -0.8 \AA  , taking the system from a half-metallic to a metallic phase.
This abrupt fall in band gap results in a sudden fall of magnetic moment of the system from 3 \bohrmag 
to 0 \bohrmag. Thus with the application of pressure, the system swaps from a highly ferromagnetic half-metallic phase 
to a non-magnetic metallic phase, which was not observed in $\mathrm{[VBz]_\infty}$ and $\mathrm{[VCp]_\infty}$.  
It is well know that the magnitude of 
the magnetic moment is strongly related to the 
lattice constant (or the volume) of the system. A possible reason for this dependence is the magneto-volume effect.\cite{magvol}
Owing to the Pauli exclusion principle for parallel spins, the electron kinetic energy of a spin-polarized state is higher,
and volume expansion relaxes the kinetic energy. Consequently, the high-spin state has a larger volume than the low-spin state. 
We here stress that the half-metallicity of $\mathrm{[VBzVCp]_\infty}$ is still conserved for stretching of upto 20\% length and
compression of upto 12\% length of the wire.       
  
\section{Conclusion}
We investigate multidecker metal-cyclopentadienyl-benzene SMWs which are derived from metal-cyclopentadienyl and metal-benzene wires. Our study finds 
that all the examined SMWs are highly stable ferromagnetic half-metals, with $\mathrm{[VBzVCp]_\infty}$ and $\mathrm{[FeBzFeCp]_\infty}$ notebly showing 
spin switching behaviour. With regard to the origin of magnetic moment in these SMWs, $E_1$ and $A_1$ bands are responsible for $\mathrm{[VBzVCp]_\infty}$, while on the 
other hand, the $E_2$ band alone contributes to the moment in $\mathrm{[FeBzFeCp]_\infty}$. 
The finite size clusters are found to exhibit an efficient spin filter property (especially BzVCpVBz) when coupled to Au electrodes on 
either side. This has been attributed to the strong resonance between minority spin states of the electrode and those of the 
molecular cluster. A close examination of $I-V_b$ characteristics for all systems clearly shows spin filtering, with Au-BzVCpVBz-Au displaying
exceptionally high performance at a bias voltage of 0.6 V. In addition to spin filtering properties, the Au-BzFeCpFeBz-Au system also show NDR. 
Additionally, the behaviour of \vbzcp when subjected to compression and stretching is also studied. The magneto-volume effect is evident as an 
abrupt change in magnetic moment on stretching and a sudden transition to a non-magnetic ordering with application of pressure.
Half-metallicity is preserved for modest amounts of stretching and compression.
Owing to better structural stability and robust electronic behaviour, we conjecture 
that (V and/or Fe)-cyclopentadienyl-benzene systems could be valuable for spintronics applications.

\section* {Acknowledgement} P.P. acknowledges CSIR, Government of India for financial support and
S.K.P. acknowledges research support from CSIR and DST, Government of India.

\end{document}